# THE USE OF GRAVITATIONAL LENSES IN THE STUDY OF DISTANT GALAXY MERGERS

A. V. Kats[1] and V. M. Kontorovich[2,3]

[1]A.Ya. Usikov Institute of Radio Physics and Electronics, NAS of Ukraine,
12, Akad. Proskura St., Kharkov, 61085, Ukraine
E-mail: ak_04@rambler.ru

[2]Institute of Radio Astronomy, National Academy of Sciences of Ukraine,
4, Chervonopraporna St., Kharkov, 61002, Ukraine
E-mail: vkont1001@yahoo.com

[3]V.N. Karazin National University of Kharkiv,
4, Svoboda Sq., Kharkov, 61022, Ukraine

Gravitational lense (GL) is currently one of the most popular astrophysical objects. They are efficiently explored for detecting the most distant galaxies (up to $z=10$ redshifts now). The first study by Pavel Victorovich Bliokh (PV) related to the subject was carried out when the wide interest to gravitational lensing did not arise. At the Institute Council meeting, some of its members where trying to convince PV not to waste time on investigations of such weak effects but rather better to devote himself to… (suggestions followed). One of the authors (V.M.K) is proud that he had strongly supported those PV activities. The future has confirmed that PV was absolutely right. The interest in GL (and the corresponding bibliography) was increased with a high speed. Here, we will only refer to the last (popular) publication by PV, where GL was dealt with [1] *[1].

PV, in his turn, had played a significant role in our studying the galaxy mergers at a time when their essential contribution was not evident yet. In particular, PV supported the graduate student report devoted to this subject at his seminar (combined with the seminar on theoretical astrophysics) in spite of the ungrounded negative attitude of some persons of influence in the Institute. His support made possible to submit the thesis to the Council for the defense of theses and defend it with honors in Moscow at the ASC LPI.

---

*[1] This wonderful little book, in which a number of exact results, including those related to gravitational lensing, has been obtained "just counted on fingers" from the physical considerations, was published in the "Quant Library" series due to the efforts of his friends and relatives already after the death of its author.

We are interested in the possibility of finding and investigation the galaxy mergers at high $z$ using GL. Mergers represent an important stage of massive galaxy evolution, when their mass function (MF) is formed along with active nuclei. As an example of the role played by GL we will refer to the observation of the galaxy merger at $z = 2.9$, cf. now-classical paper by Borys, et al., [2], along with Berciano Alba, et al. [3] comments.

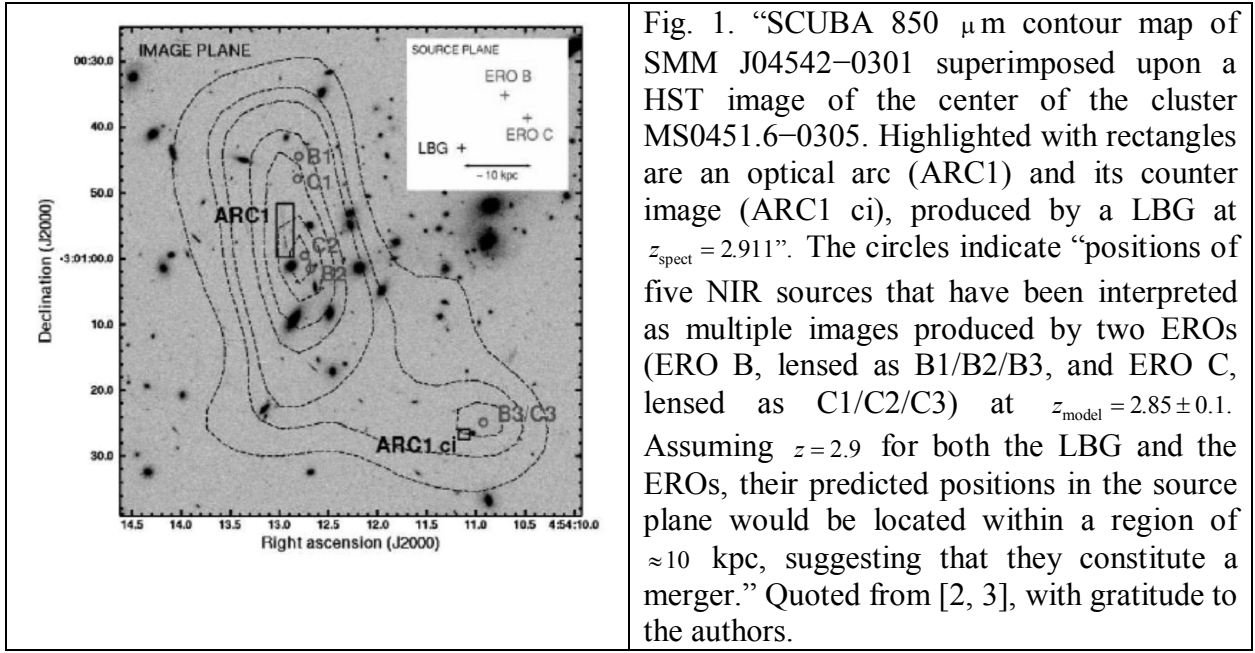

Fig. 1. "SCUBA 850 μm contour map of SMM J04542−0301 superimposed upon a HST image of the center of the cluster MS0451.6−0305. Highlighted with rectangles are an optical arc (ARC1) and its counter image (ARC1 ci), produced by a LBG at $z_{spect} = 2.911$". The circles indicate "positions of five NIR sources that have been interpreted as multiple images produced by two EROs (ERO B, lensed as B1/B2/B3, and ERO C, lensed as C1/C2/C3) at $z_{model} = 2.85 \pm 0.1$. Assuming $z = 2.9$ for both the LBG and the EROs, their predicted positions in the source plane would be located within a region of $\approx 10$ kpc, suggesting that they constitute a merger." Quoted from [2, 3], with gratitude to the authors.

Here, the following abbreviations are used: sub-mm source, SMM; Lyman Break Galaxy, LBG; sub-mm galaxy, SMG; extremely red object, ERO. Due to the radio-FIR correlation in star-forming galaxies, the radio interferometric observations were used to obtain a high-resolution rest-frame far-IR emission that observed in the sub-mm. (Here the astronomy scale of EM spectrum is used, i.e., FIR upper wavelength is about 300 μm.) As a result, the ERO pair and the LBG may constitute a merger at $z = 2.9$ (see the merger model in [2]).

The merger of distant galaxies ($z \approx 3$) [2, 3], as actually observed phenomenon, will serve us as a starting point for the analysis evolution of the galaxy mass function.

Below we study the explosive galaxy evolution resulting from the merger process with a low mass increase (minor mergers) assuming that along with the low-mass background, there exists a source of high-mass galaxies (the ones, segregating from the general expansion). Note that the resulting MF possesses power-law asymptote with the exponent $\alpha$ coinciding with the Shechter index[*2], 1.25 at $z = 0$.

Consider solutions of the Smoluchowski kinetic equation (KE) in the differential form supposing that the main contribution is due to mergers of the low-mass galaxies with the massive ones with the corresponding merging probability, $U(M_1, M_2) \simeq 0.5 C M_1^u$ for $M_2 \ll M_1$. The main contribution to the collision integral follows from small masses of order $M_*$ and less:

$$\frac{\partial}{\partial t} f(M, t) + C\Pi \frac{\partial}{\partial M}\left[M^u f(M, t)\right] = \phi(M, t), \qquad (1)$$

$$\Pi = \Pi(t) = \int dM_2 M_2 f(M_2, t),$$

where $\Pi$ is approximately the total mass of low-mass galaxies. Rewriting Eq. (1) as

$$\frac{\partial}{\partial t} F(M, t) + C\Pi M^u \frac{\partial}{\partial M} F(M, t) = \Phi(M, t),$$

$$F(M, t) = M^u f(M, t), \qquad (2)$$

$$\Phi(M, t) = M^u \phi(M, t),$$

and using the method of characteristics we arrive at the following system of ordinary differential equations (ODEs),

$$dM/dt = C\Pi M^u,$$

$$dF/dt = \Phi.$$

It is easy to find one of the first integrals, $a(M, t)$, by strict integration[*3] of the first relation:

$$\tau(t) - \frac{M^{1-u}}{1-u} = a(M, t) = \text{const},$$

$$\tau(t) \equiv C \int_0^t dt \Pi(t). \qquad (3)$$

For $\Phi \neq 0$, the second independent integral of the system is to be obtained from the second ODE. The mass variable, $M$, here is the solution of the first ODE and thus it is to be treated as time-dependent.

---

[*2] The Shechter index is often defined with the sign reversed to here chosen.
[*3] We consider parameters C and $u$, determined by the probability of mergers, being time-independent, thus suggesting fairly rapid evolution of the explosive solutions of KE.

Namely,

$$M = \mu(a, t), \quad \mu(a, t) = \left[(u-1)(a - \tau(t))\right]^{1/1-u},$$

where $a$ is the integration constant. In order to solve this equation we have to specify the source term, $\Phi(M, t)$. Let us restrict ourselves with a localized source, i. e., suppose that $\Phi(M, t) = \delta(M - \bar{M}(t))\Phi(t)$, where $\Phi(t)$ is some time-dependent function. Such representation simulates the mass $\bar{M}(t)$ separation from the global expansion at the moment $t$. Integrating now the second ODE, we obtain the following independent first integral of the ODEs system,

$$F - K(a, t) = b(M, t) = \text{const},$$

where

$$K(a, t) = \int_0^t dt \, \delta\left[\mu(a, t) - \bar{M}(t)\right] \Phi(t) =$$

$$= \sum_n \Phi(t_n) \theta(t - t_n) \left| \frac{d}{dt}\left[\mu(a, t) - \bar{M}(t)\right]\right|_{t=t_n}^{-1}, \qquad (4)$$

and $t_n$ denotes the roots of equation

$$\mu(a, t) - \bar{M}(t) = 0.$$

Using the initial condition,

$$f(M, 0) - \bar{M}(t) = 0,$$

we find the KE solution:

$$f(M, t) = f_s(M, t) + f_{in}(M, t),$$

$$f_s(M, t) = M^{-u} K\left(\tau + \frac{M^{1-u}}{u-1}, t\right),$$

$$f_{in}(M, t) = \left[(u-1)\tau M^{u-1} + 1\right]^{\frac{u}{1-u}} f_0\left\{M\left[(u-1)\tau M^{u-1} + 1\right]^{\frac{1}{1-u}}\right\}. \qquad (5)$$

In order to specify the source-induced contribution to the mass function, $f_s(M, t)$, we need to derive an explicit expression for the function $K(a, t)$, cf. Eq. (4), i. e. to find the corresponding roots. For the simplest case, $u = 2$, $\bar{M}(t) = t/A$, $\Pi(t) = \Pi = \text{const}$, we have

$\mu(a,t) = (a - C\Pi t)^{-1}$, and the equation for the roots becomes $C\Pi t^2 - at + A$. Consequently, real roots exist under the condition, $a \geq a_{cr} \equiv 2\sqrt{AC\Pi}$, and in terms of the normalized quantities, $T \equiv t\sqrt{C\Pi/A}$, $\tilde{a} \equiv a/a_{cr}$, they are $T_\pm = \tilde{a} \pm \sqrt{\tilde{a}^2 - 1}$. Then the source-induced term of the MF becomes

$$M^u f_s(M,t) = K\left(\tau + \frac{1}{M}, t\right) = \frac{A}{2\sqrt{\tilde{a}^2(M,t) - 1}} \sum_\pm \frac{\Phi(t_\pm(M,t))}{|T_\pm(M,t)|} \theta(\tilde{a}(M,t) - 1)\theta(T - T_\pm(M,t)). \quad (6)$$

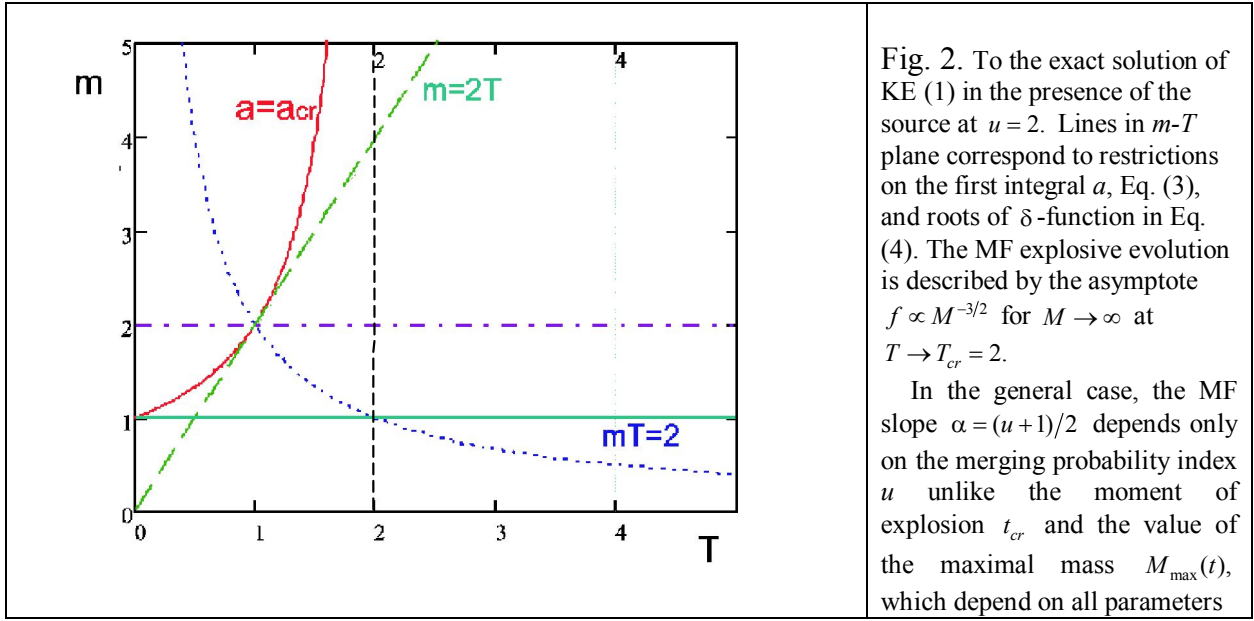

Fig. 2. To the exact solution of KE (1) in the presence of the source at $u = 2$. Lines in $m$-$T$ plane correspond to restrictions on the first integral $a$, Eq. (3), and roots of $\delta$-function in Eq. (4). The MF explosive evolution is described by the asymptote $f \propto M^{-3/2}$ for $M \to \infty$ at $T \to T_{cr} = 2$.

In the general case, the MF slope $\alpha = (u+1)/2$ depends only on the merging probability index $u$ unlike the moment of explosion $t_{cr}$ and the value of the maximal mass $M_{max}(t)$, which depend on all parameters

It is convenient to introduce the normalized mass, $m$, $m = a_{cr}M = 2\sqrt{AC\Pi} \cdot M$. Then $\tilde{a}(M,t) = T/2 + m^{-1}$ and it becomes evident that we arrive at the MF explosive evolution: the MF expands to the infinite mass region at finite time, $T \to T_{cr} = 2$, i. e., $t \to t_{cr} = \sqrt{A/(C\Pi)}$. The asymptotic behavior of the exact solution obtained with such a source term is of the power-law type: within the region $1 \ll m \ll (1 - T/2)^{-1} \equiv m_{max}(t)$ we obtain $f(M,t) \propto M^{-3/2}$. In the general case, $u > 1$, we also have the explosive evolution with power-law asymptotic at high masses "$M \to \infty$"

$$f(M, t \to t_{cr}) \propto M^{-(u+1)/2}.$$

For calculating the index homogeneity of merger probability $u$ we use the Tally-Fisher or Faber-Jackson laws, linking the mass of galaxy and its radius. For large masses we take into account gravitational focusing in the merger cross-section [4].

The observed growth of the MF power index as $z$ increases (up to $\alpha \approx 2$ at $z = 6$ [5]) can result from the evolutionary change of the merger mechanisms. Indeed, the steepest MF can arise due to the evolution of the initial distribution $f_{in}$ (5) with $\alpha = u = 2$ for large $z$ values and for relatively small masses. At lower $z$ and larger masses, the gravitational focusing in the cross-section results in $\alpha = u = 1.5$. The source-governed term of MF (5), $f_s$, results in $\alpha = 1.5$ for $u = 2$ (small masses), and $\alpha = 1.25$ for $u = 1.5$ (large masses, gravitational focusing). The latter corresponds to $z = 0$ and agrees with the well-known from observations value.

More detail description and evaluation of maximal masses see in [6].

## Conclusions

The galaxy merger process via the gravitational interaction possesses the "explosive character" due to the merging probability dependence on the galaxy masses such that the probability increases with mass faster than its first power. As a result, there arises the critical time moment, which may correspond to the epoch of the massive galaxies formation (see refs in [7-9]). The gravitational lensed observations of galaxy mergings offer great opportunities for approval of this process. Small galaxy mergers can explain the observed in the Hubble ultra deep field [5] evolution of the slope of galaxy luminosity function. This is the main concrete result of this work.